\documentclass[11pt,final]{article}

\def\xetex{1}

\ifx\xetex\undefined
\usepackage[utf8]{inputenc}
\usepackage[T1]{fontenc}
\usepackage{lmodern} 
\fi

\usepackage{cite}
\usepackage{array}
\usepackage{url}
\usepackage[english]{babel}
\usepackage{algorithm}
\usepackage{algorithmic}
\usepackage{tikz-cd}
\usepackage{microtype}
\usepackage{enumerate}
\usepackage{amsmath}
\usepackage{booktabs}
\usepackage{cite}
\usepackage{varioref}
\usepackage[pdfpagelabels=true]{hyperref}
\usepackage{cleveref}
\usepackage{verbatim}
\usepackage{listings}
\usepackage{authblk}

\hypersetup{
    linktoc = page,
    pdfpagemode = UseNone,
    colorlinks,
    linkcolor={red!50!black},
    citecolor={blue!50!black},
    urlcolor={blue!80!black}
}

\lstdefinelanguage
[aarch64]{Assembler}     
[x86masm]{Assembler}
{morekeywords=
	{adds,tbnz,ldrb,strb,subs,tbz,ands,adr,eon,lsr,
	 bics,ldr,bfm,ubfm,orr,eor,ccmp,cbz,cbnz,b.cond,
	 ldpsw,ldnp,ldp,ldrh,ldurb,ldxrh,ldtrb,ldtrh,ldurh,
	 stnp,stp,strh,cmhi,cmgt,umin,smin,smax, umax, 
	 usubw2, ushl, srshl, sqshl, urshl, uqshl,
	 sshl, ssubw2, rsubhn2, sqdmlal2, subhn2,
	 umlsl2, smlsl2, uabdl2, sabdl2, sqdmlsl2,
	 fcvtxn2, fcvtn2, raddhn2, addhn2, fcvtl2, 
	 uqxtn2, sqxtn2, uabal2, sabal2, sri, sli,
	 uabd, sabd, ursra, srsra, uaddlv, saddlv,
	 sqshlu, shll2, zip2, zip1, uzp2, mls, trn2},
    morecomment=[s]{/*}{*/}}

\newcommand{\ignore}[1]{ }
\newcommand{\hex}[1]{\texttt{0x#1}}
\newcommand{\bin}[1]{\texttt{0b#1}}

\begin{document}

\title{ARMv8 Shellcodes from `A' to `Z'}

\author{Hadrien Barral}
\author{Houda Ferradi}
\author{R\'emi G\'eraud}
\author{Georges-Axel Jaloyan}
\author{David Naccache}
\affil{DIENS, \'Ecole normale sup\'erieure, \\CNRS, PSL University, Paris, France}

\renewcommand\Authands{ and }


\maketitle

\begin{abstract}
We describe a methodology to automatically turn arbitrary ARMv8 programs into alphanumeric executable polymorphic shellcodes.
Shellcodes generated in this way can evade detection and bypass filters, broadening the attack surface of ARM-powered devices such
as smartphones. 
\end{abstract}

\section{Introduction}

Much effort has been undertaken in recent years to secure smartphones and tablets. For such devices, software security is a challenge: on the one hand, most software applications are now developed by third-parties; on the other hand, defenders are restrained as to which watchdogs to install, and how efficient they can be, given these devices' restricted computational capabilities and limited battery life.

In particular, it is important to understand how countermeasures fare against one of the most common security concerns: memory safety issues. Using traditional buffer overflow exploitation techniques, an attacker may exploit a vulnerability to successfully execute arbitrary code, and take control of the device~\cite{PhrackBOF}. The relatively weak physical defenses of mobile devices tend to make memory attacks a rather reliable operation~\cite{davi2011privilege}. 

In a typical \emph{arbitrary code execution} (ACE) scenario, the attacker can run a relatively short self-contained program of their choosing. It is called a \emph{shellcode} -- as it enables an opponent to gain control of the device by opening a shell, or alter the memory regardless of the security policy. Such shellcodes are easy to distribute and weaponize, as shown by many off-the-shelf shellcodes available within exploitation frameworks such as Metasploit~\cite{Metasploit}.

To launch the attack, the opponent sends the shellcode to a vulnerable application, either by direct input, or via a remote client. However, before doing so the attacker might have to overcome a number of difficulties: if the device has a limited keyboard for instance, some characters might be hard or impossible to type; or filters may restrict the available character set of remote requests for instance. A well-known situation where this happens is input forms on web pages, where input validation and escaping is performed by the server.

This paper describes an approach allowing to compile arbitrary shellcodes into executable code formed from a very limited subset of the ASCII characters. We focus on \emph{alphanumeric} shellcodes, and target the ARM-v8A architecture, to illustrate our technique. More specifically, we will work with the AArch64 instruction set, which powers the Exynos 7420 (Samsung Galaxy S6), Project Denver (Nexus 9), ARM Cortex A53 (Raspberry Pi 3), A57 (Snapdragon 810), A72, Qualcomm Kryo (Snapdragon 818, 820 and 823), as well as the Apple A7 and A8 ARMv8-compatible cores (Apple iPhone 5S/6).

\subsection{Prior and related work}

The idea to write alphanumeric executable code first stemmed as a response to anti-virus or hardening technologies that were based on the misconception that executable code is not ASCII-printable.
Eller~\cite{Eller} described the first ASCII-printable shellcode for Intel platforms able to bypass primitive buffer-overflow protection techniques. This was shortly followed by RIX~\cite{PhrackIA32} on the IA32 architecture. Later, Mason \emph{et al.}~\cite{DBLP:conf/ccs/MasonSMM09} showed a technique to automatically turn IA32 shellcodes into English shellcodes, statistically indistinguishable from any other English text. Obscou~\cite{PhrackUnicode} managed to obtain Unicode-proof shellcodes that work despite the limitation that no zero-character can appear in the middle of a standard C string. All the above constructions are relying on existing shellcode writing approaches and require manual fine-tuning.

Basu \emph{et al.}~\cite{Basu} developed an algorithm for automated shellcode generation targeting the x86 architecture. The Metasploit project provides the \texttt{msfvenom} utility, which can turn arbitrary x86 programs into alphanumeric x86 code. Both \texttt{UPX}\footnote{See \url{http://upx.sf.net}.} and \texttt{msfvenom} can generate self-decrypting ARM executables, yet neither provide alphanumeric encodings for this platform.

More recently, Younan \emph{et al.} generated alphanumeric shellcodes for the ARMv5 architecture~\cite{PhrackARM,Younan}. They provide a proof that the subset of alphanumeric commands is Turing-complete, 
by translating all Brainfuck~\cite{raiter07,faase07,cristofani07} commands into alphanumeric ARM code snippets. 

\subsection{Our contribution}

This paper describes, to the best of the authors' knowledge, the first program turning \emph{arbitrary} ARMv8 code into alphanumeric executable code. The technique is generic and may well apply to other architectures.
Besides solving a technical challenge, our tools produce valid shellcodes that can be used to try and take control of a device.

Our global approach is the following: we first identify a subset $\Sigma$ of minimal Turing-complete alphanumeric instructions, and use $\Sigma$ to write an in-memory decoder. The payload is encoded offline (with an algorithm that only outputs alphanumeric characters), and is integrated into the decoder. The whole package is therefore an alphanumeric program, and allows ACE. All source files are provided in the appendices.

\section{Preliminaries}

\subsection{Notations and definitions}

Throughout this paper, a string will be defined as \emph{alphanumeric} if it only contains upper-case or lower-case letters of the English alphabet, and numbers from 0 to 9 included. When writing alphanumeric code, spaces and return characters are added for reading convenience but are not part of the actual code. Words are 32-bit long. We call \emph{polymorphic}, a code that can be mutated into another one with the same semantics. This mutation is performed by another program called the \emph{polymorphic engine}.

When dealing with numbers we use the following convention: plain numbers are in base 10, 
numbers prefixed by \hex{} are in hexadecimal format, and numbers prefixed by \bin{} are in binary format. 
The little-endian convention is used throughout this paper for alphanumeric code, to remain consistent with ARMv8 internals. However, registers will be considered as double-words or words; each 32-bit register $\texttt{W} = \texttt{W}_\text{high} \texttt{W}_\text{low}$ is split into a most significant 16 bits half-word $\texttt{W}_\text{high}$ and a least significant 16 bits half-word $\texttt{W}_\text{low}$.

$S[i]$ denotes $i$-th byte of a string $S$. Each byte is 8 bits long.

\subsection{Vulnerable applications and platforms}

To perform a buffer overflow attack, we assume there exists a vulnerable application on the target device.
Smartphone applications are good candidates because (1) they can easily be written in languages that do not check
array bounds; and (2) they can be spread to many users via application marketplaces.

Note that on Android platforms, applications are often written in Java which implements implicit bound checking.
At first glance it may seem that this protects Java applications from buffer overflow attacks.
However, it is possible to access C/C++ code libraries via the Java Native Interface (JNI), for performance reasons. Such vulnerabilities were exposed in the JDK~\cite{tan2008empirical}.

\subsection{ARMv8 AArch64}

AArch64 is a new ARM-v8A instruction set. AArch64 features 32-bit naturally aligned instructions. There are 32 general purpose 64-bit registers \texttt{Xi} ($0 \leq i < 32$) and 32 floating-point registers. The 32 least significant bits (LSB) of each \texttt{Xi} is denoted by \texttt{Wi}, that can be used directly in many instructions. Younan \emph{et al.}~\cite{Younan} use the fact that in AArch32 (32-bits ARM architecture), almost all instructions can be executed conditionally via a condition code checked against the \texttt{CPSR} register. In AArch64, this is not the case anymore. Only specific instructions, such as branches, can be made conditional: this renders their approach nugatory.

Each instruction is composed of an \texttt{opcode} and zero or more \texttt{operands}. The \texttt{opcode} specifies the operation to perform while the \texttt{operands} consist in addresses, register numbers, or constants being manipulated. As an example, the instruction:
\begin{lstlisting}
    ldr    x16, PC+0x60604
\end{lstlisting}
is assembled as \hex{58303030}\footnote{Which is \texttt{01011000001100000011000000110000} in binary. Incidentally, this instruction is alphanumeric and corresponds to the ASCII string \texttt{000X}. Note the little endianness of the string.} and decoded as follows~\cite{ARMv8Ap536}:

{\footnotesize\ttfamily
\begin{center}
\begin{tabular}{l|l|l|l|l|l}
0 1  & 0  1  1  & 0  & 0  0  & {imm19} & {Xt} \\
\end{tabular}
\end{center}
}

Bits 0 to 4 encode the reference number of the 64-bit register \texttt{Xt}. Bits 5 to 23 encode the load relative offset counted in words, encoded as the immediate value \texttt{imm19}.

An interesting feature is that immediate values and registers often follow each other in instructions, as shown above with \texttt{imm19} and \texttt{Xt}. This is a real advantage for creating alphanumeric shellcodes, as it indicates that
instructions who share a prefix are probably related. For instance \texttt{000X} and \texttt{100X} turn out to decode respectively into
\begin{lstlisting}
    ldr x16, PC+0x60604
\end{lstlisting}
and
\begin{lstlisting}
    ldr x17, PC+0x60604
\end{lstlisting}
Thus it is relatively easy to modify the operands of an existing instruction.

\subsection{Shellcodes}

A \emph{shellcode} is a set of machine code instructions injected into a running program. 
To that end, an attacker would for instance exploit a buffer overflow vulnerability by inserting executable code into the stack, and control the current stack frame's return address.
As a result, when the target program's current function returns, the attacker's code gets executed. Other strategies might be employed to achieve that goal, which are not within the scope of this study.

It is common practice to flood the buffer with a \emph{nopsled}, \emph{i.e.} a sequence of useless operations, which has the added benefit of allowing some imprecision in the return address. 

Shellcodes may execute directly, or employ some form of evasion strategy such as filter evasion, encryption or polymorphism. The latter allows having a large number of different shellcodes that have the same effect, which decreases their traceability. In these cases the payload must be encoded in a specific way, and must decode itself at runtime.

In this work, we encode the payload in a filter-friendly way and equip it with a decoder (or \emph{vector}). The vector \emph{itself} must be filter-friendly, and is usually handwritten. Hence designing a shellcode is a tricky art. 

\section{Building the instruction set}

In order to build the alphanumeric subset of AArch64, we generated all 14,776,336 alphanumeric 32-bit words using the custom-made program provided in Appendix~\ref{app:program1}. For each 4 bytes values obtained, we tentatively disassembled it using \texttt{objdump}\footnote{We used the options \texttt{-D -{}-architecture aarch64 -{}-target binary}.} for the AArch64 architecture, in the hope that these words correspond to valid and interesting instructions.

For instance, the word \texttt{000X} corresponds to an \texttt{ldr} instruction, whereas the word \texttt{000S} does not correspond to any valid AArch64 instruction:
\begin{lstlisting}
    58303030 ldr    x16, PC+0x60604
    53303030 .inst  0x53303030 ; undefined
\end{lstlisting}

Alphanumeric words that do not correspond to any valid instruction (``undefined'') are removed from our set. Valid instructions are finally classified as pertaining to data processing, branch, load/store, etc. At this step we established a first list $\mathcal A_0$ of all valid alphanumeric AArch64 instructions.

From $\mathcal A_0$, we constructed a set $\mathcal A_1$ of opcodes for which there exists \emph{at least one} operand instance making it alphanumeric. $\mathcal A_1$ is given in Appendix~\ref{app:list1}. Finally, we extracted from $\mathcal A_1$ the instructions which could be used to prototype higher-level constructs. This final list is called $\mathcal A_\text{max}$.

\subsection{Data processing}
The following data processing instructions belong to $\mathcal A_{\rm max}$:
\begin{lstlisting}
    adds (immediate) 32-bit
    sub  (immediate) 32-bit
    subs (immediate) 32-bit
    bfm  32-bit
    ubfm 32-bit
    orr  (immediate) 32-bit
    eor  (immediate) 32-bit
    ands (immediate) 32-bit
    adr
    sub  32 extended reg
    subs 32 extended reg
    sub  32 shifted reg
    subs 32 shifted reg
    ccmp (immediate)
    ccmp (register)
    eor  (shifted register) 32-bit
    eon  (shifted register) 32-bit
    ands (shifted register) 32-bit
    bics (shifted register) 32-bit
\end{lstlisting}
Furthermore, we are constrained by $\mathcal A_{\rm max}$ to have the \texttt{sf} bit of each instruction set to $0$, which restricts us to only the 32-bit variant of most instructions, hindering our ease to modify the upper 32 bits of a register.

\subsection{Branches}
Only the following conditional jumps are available in $\mathcal A_{\rm max}$:
\begin{lstlisting}
    cbz  32-bit
    cbnz 32-bit
    b.cond
    tbz
    tbnz
\end{lstlisting}
It is quite easy to turn a conditional jump into a non-conditional jump. However, only \texttt{tbz} and its opposite \texttt{tbnz} have a realistic use for loops, as the three other branching instructions require an offset too large to be useful. Therefore, $\mathcal A_{\rm max}$ contains only \texttt{tbz} and \texttt{tbnz} as branching instructions.

\subsection{Exceptions and system}
Neither exceptions nor system instructions are available. This means that we cannot use syscalls, nor clear the instruction or data cache. This makes writing higher-level code challenging and implementation-dependent.
 
\subsection{Load and stores}
Many load and stores instructions can be alphanumeric. This requires fine tuning to achieve the desired result, as limitations on the various load and store instructions are not consistent across registers.
\subsection{SIMD, floating point and crypto}
No floating point or cryptographic instruction is alphanumeric. Some \emph{Single Instruction on Multiple Data} (SIMD) are available, but the instructions moving data between SIMD and general purposes registers are not alphanumeric. This limits the use of such instructions to very specific cases. Therefore, we did not include any of these instructions in $\mathcal A_{\rm max}$.

\section{High-level constructs}

\label{sec:secondlook}
A real-world program may need information about the state of registers and memory, including the program counter and processor flags. This information is not immediately obtainable using $\mathcal A_{\rm max}$. We overcome this difficulty by providing higher-level constructs, which can then be combined to form more complex programs. Indeed it turns out that $\mathcal A_\text{max}$ is Turing-complete. Those higher-level constructs also make it easier to turn a program polymorphic, by just providing several variants of each construct.

\subsection{Registers operations}

\subsubsection{Zeroing a register}
\label{sec:zeroingregister}
There are multiple ways of setting an AArch64 register to zero. One of them which is alphanumeric and works well on many registers consists in using two \texttt{and} instructions with shifted registers. However, we only manage to reset the register's 32 LSBs. This becomes an issue when dealing with addresses for instance.

As an example, the following instructions reset the 32 LSBs of $\texttt{x17}$, denoted by $\texttt{w17}$:
\begin{lstlisting}
    ands w17, w17, w17, lsr #16
    ands w17, w17, w17, lsr #16
\end{lstlisting}
This corresponds to the alphanumeric code \texttt{1BQj1BQj}. The following table summarizes some of the zeroing operations we can perform:
\begin{center}
    \ttfamily
\begin{tabular}{lll}
$a$ & $a_\text{low} \gets 0$ & lsr \\
w2 & BlBjBlBj &  27\\
w3 & cdCjcdCj &  25\\
w10 & JAJjJAJj & 16\\
w11 & kAKjkAKj & 16\\
w17 & 1BQj1BQj & 16\\
w18 & RBRjRBRj & 16\\
w19 & sBSjsBSj & 16\\
w25 & 9CYj9CYj & 16\\
w26 & ZCZjZCZj & 16\\
\end{tabular}
\end{center}

\subsubsection{Loading arbitrary values into a register}
\label{sec:incrdecr}
Loading a value into a register is the cornerstone of any program. Unfortunately there is no
direct way to perform a load using only alphanumeric instructions. We hence opted for an indirect strategy using a sequence of \texttt{adds} and \texttt{subs} instructions with different immediates, changing the value of the registers to the desired amount. 
One of the constraints is that this immediate constant must be quite large for the instruction to be alphanumeric. 
In particular, we selected two consecutive constants for increasing and decreasing registers, using an \texttt{adds/subs} pair. By repeating such operations we can set registers to arbitrary values.

For instance, to add 1 to the register \texttt{w11} we can use:
\begin{lstlisting}
    adds    w11, w11, #0xc1a
    subs    w11, w11, #0xc19
\end{lstlisting}
which is encoded by \texttt{ki01ke0q}. And similarly to subtract~1:
\begin{lstlisting}
    subs    w11, w11, #0xc1a
    adds    w11, w11, #0xc19
\end{lstlisting}
which is encoded by \texttt{ki0qke01}.

The following table summarizes the available increment and decrement operations:
\begin{center}
    \ttfamily
\begin{tabular}{lll}
$a$ & $a \gets a+1$ & $a \gets a-1$\\
w2  & Bh01Bd0q & Bh0qBd01 \\
w3  & ch01cd0q & ch0qcd01 \\
w10 & Ji01Je0q & Ji0qJe01 \\
w11 & ki01ke0q & ki0qke01 \\
w17 & 1j011f0q & 1j0q1f01 \\
w18 & Rj01Rf0q & Rj0qRf01 \\
w19 & sj01sf0q & sj0qsf01 \\
w25 & 9k019g0q & 9k0q9g01 \\ 
w26 & Zk01Zg0q & Zk0qZg01 \\
\end{tabular}
\end{center}
In this paper, we manually selected registers and constants to achieve the desired value. However, it would be much more efficient to solve a knapsack problem, if one were to do this at a larger scale. As we will see later on, the values above are sufficient for our needs.

\subsubsection{Moving a register}
Moving a register $A$ into $B$ can be performed in two steps: first we set the destination register to zero, and then we \texttt{xor} it with the source register. The \texttt{xor} operation is described in \Cref{sec:xor}.

An \textit{ad hoc} method we used for moving \texttt{w11} into \texttt{w16} is:
\begin{lstlisting}
    adds w17, w11, #0xc10
    subs w16, w17, #0xc10
\end{lstlisting}
which is encoded by \texttt{qA010B0q}. We will later use this approach when designing a logical \texttt{and} operation.
\subsection{Bitwise operations}
\subsubsection{Exclusive OR}
\label{sec:xor}

The \texttt{xor} operation $B \gets A \oplus B$ can be performed as follows: we split the two input registers into their higher and lower half-words, and use a temporary register $C$.
\begin{align*}
C & \gets 0 \\
C_\text{high} & \gets C_\text{high} \oplus \neg{A_\text{low}} \\
C_\text{low} & \gets   C_\text{low} \oplus \neg{A_\text{high}} \\
B_\text{high} & \gets  B_\text{high} \oplus \neg{C_\text{low}}  =  B_\text{high} \oplus {A_\text{high}} \\
B_\text{low} & \gets  B_\text{low} \oplus \neg{C_\text{high}}  =  B_\text{low} \oplus {A_\text{low}}
\end{align*}
This gives the following code: 
\begin{lstlisting}
    eor (xor) b:= a eor b, 
      c = w17 a = w16-25 b= w18-25
    c:=0
    eon c c a lsl 16
    eon c c a lsr 16
    eon b b c lsl 16
    eon b b c lsr 16
\end{lstlisting}

In particular, when \texttt{c} = \texttt{w17}, the following instructions can be used: 
\begin{center}
    \ttfamily
\begin{tabular}{lll}
$a$    & $b$   & $b \gets a \oplus b$   \\
w16  & w16 & 1B0J1BpJRB1JRBqJ \\
w16  & w18 & 1B0J1BpJRB1JRBqJ \\
w16  & w19 & 1B0J1BpJsB1JsBqJ \\
w16  & w25 & 1B0J1BpJ9C1J9CqJ \\
w16  & w26 & 1B0J1BpJZC1JZCqJ \\
w18  & w19 & 1B2J1BrJsB1JsBqJ \\
w18  & w25 & 1B2J1BrJ9C1J9CqJ \\
w18  & w26 & 1B2J1BrJZC1JZCqJ \\
w19  & w25 & 1B3J1BsJ9C1J9CqJ \\
w19  & w26 & 1B3J1BsJZC1JZCqJ \\
w20  & w25 & 1B4J1BtJ9C1J9CqJ \\
w20  & w26 & 1B4J1BtJZC1JZCqJ \\
w21  & w25 & 1B5J1BuJ9C1J9CqJ \\
w21  & w26 & 1B5J1BuJZC1JZCqJ \\
w22  & w25 & 1B6J1BvJ9C1J9CqJ \\
w22  & w26 & 1B6J1BvJZC1JZCqJ \\
w23  & w25 & 1B7J1BwJ9C1J9CqJ \\
w23  & w26 & 1B7J1BwJZC1JZCqJ \\
w24  & w25 & 1B8J1BxJ9C1J9CqJ \\
w24  & w26 & 1B8J1BxJZC1JZCqJ \\
w25  & w26 & 1B9J1ByJZC1JZCqJ \\
\end{tabular}
\end{center}

\subsubsection{Logical NOT}
We use the fact that $\neg b = b \oplus (-1)$ which relies on negative number being represented in the 
usual two's complement format. Thus we can use the operations described previously:
\begin{align*}
C & \gets 0 \\
C & \gets C-1  \\
B & \gets C \oplus B \\
\end{align*}
\subsubsection{Logical AND}
\label{sec:AND}

The \texttt{and} operation is more intricate and requires three temporary registers $C$, $E$, and $F$. We manage to do it by anding the lower and the upper parts of the two operands into a third register
as follows:
\begin{align*}
D & \gets  0 \\
C & \gets  0 \\
E & \gets  0 \\
F & \gets  0 \\
C_\text{high} & \gets  C_\text{high} \oplus \neg{B_\text{low}} \\
E_\text{high} & \gets  E_\text{high} \oplus \neg{A_{low}} \\
F_\text{low} & \gets  F_\text{low} \oplus \neg{E_\text{high}} =  {A_\text{low}} \\
D_\text{low} & \gets  F_\text{low} \wedge \neg{C_\text{high}} =  {A_\text{low}} \wedge B_\text{low} \\
C & \gets  0 \\
E & \gets  0 \\
F & \gets  0 \\
C_\text{low} & \gets  C_\text{low} \oplus \neg{B_\text{high}} \\
E_\text{low} & \gets  E_\text{low} \oplus \neg{A_\text{high}} \\
F_\text{high} & \gets  F_\text{high} \oplus \neg{E_\text{high}} = {A_\text{high}} \\
D_\text{high} & \gets  F_\text{high} \wedge \neg{C_\text{low}} = {A_\text{high}} \wedge B_\text{high} \\
\end{align*}
Which corresponds to the assembly code: 
\begin{lstlisting}
    and: d:= a and b
    c,d,e,f:=0
    eon c c b lsl 16
    eon e e a lsl 16
    eon f f e lsr 16
    bics d f c lsr 16
    c,e,f:=0
    eon c c b lsr 16
    eon e e a lsr 16
    eon f f e lsl 16
    bics d f c lsl 16
\end{lstlisting}

As an illustration of this technique, let
\begin{align*}
A & \gets \texttt{w18}, \quad B \gets \texttt{w25}, \quad C \gets\texttt{w17}, \\
D & \gets \texttt{w11}, \quad E \gets \texttt{w19}, \quad F \gets \texttt{w26}
\end{align*}
which corresponds to computing $\texttt{w11} \gets \texttt{w18} \wedge \texttt{w25}$. This
gives the following assembly code:
\begin{lstlisting}
    ands    w11, w11, w11, lsr #16
    ands    w11, w11, w11, lsr #16
    ands    w17, w17, w17, lsr #16
    ands    w17, w17, w17, lsr #16
    ands    w19, w19, w19, lsr #16
    ands    w19, w19, w19, lsr #16
    ands    w26, w26, w26, lsr #16
    ands    w26, w26, w26, lsr #16
    eon     w17, w17, w25, lsl #16
    eon     w19, w19, w18, lsl #16
    eon     w26, w26, w19, lsr #16
    bics    w11, w26, w17, lsr #16
    ands    w17, w17, w17, lsr #16
    ands    w17, w17, w17, lsr #16
    ands    w19, w19, w19, lsr #16
    ands    w19, w19, w19, lsr #16
    ands    w26, w26, w26, lsr #16
    ands    w26, w26, w26, lsr #16
    eon     w17, w17, w25, lsr #16
    eon     w19, w19, w18, lsr #16
    eon     w26, w26, w19, lsl #16
    bics    w11, w26, w17, lsl #16
\end{lstlisting}
This gives the following alphanumeric sequence of instructions:
\begin{lstlisting}
kAKjkAKj1BQj1BQjsBSjsBSjZCZjZCZj1B9JsB2J
ZCsJKCqj1BQj1BQjsBSjsBSjZCZjZCZj1ByJsBrJ
ZC3JKC1j
\end{lstlisting}
We provide in Appendix~\ref{app:codeAND} a program generating more sequences of this type.

\subsection{Load and store operations}
There are several load and store instructions available in $\mathcal A_\text{max}$. We will only focus or \texttt{ldrb} (which loads a byte into a register) and \texttt{strb} (which stores the lowest byte of a register into memory).

\texttt{ldrb} is available with the basic addressing mode: \texttt{ldrb wA, [xP, \#n]} which loads the byte at address \texttt{xP+n} into \texttt{wA}. Furthermore, we must use a value of \texttt{n} that makes the whole instruction alphanumeric, but this is not a truly limiting constraint. Besides, we can chain different values of \texttt{n} to load consecutive bytes from memory without modifying \texttt{xP}. 

Another addressing mode which can be used is \texttt{ldrb wA, [xP, wQ, uxtx]}. This will extend the 32-bits register \texttt{wQ} into a 64 bit one, padding the high bits with zeros, which removes the need for an offset.

As an illustration, we load a byte from the address pointed by \texttt{x10} and store it to the address pointed by \texttt{x11}. First, we initialize a temporary register to zero and remove the \texttt{ldrb} offset from \texttt{x10} using the previous constructs.
\begin{align*}
    \texttt{w19} & \gets 0 \\
    \texttt{w25} & \gets \texttt{w25} - 77
\end{align*}
Then, we can actually load and store the byte using the two following instructions corresponding to the alphanumeric executable code \texttt{Y5A9yI38}.
\begin{lstlisting}
    ldrb    w25, [x10, #77]
    strb    w25, [x11, w19, uxtw]
\end{lstlisting}

\subsection{Pointer arithmetic}
As mentioned previously we only control the 32 LSBs of \texttt{XP} with data processing instructions. Thus, if addresses are in the 4\,GB range, we can use the data-instructions seen previously to add 1, load the next byte, and loop on it. If the addresses do not fit into 32 bits, any use of data instructions will clear the 32 upper bits. Thus, we need a different approach. 

We use another addressing mode which reads a byte from the source register, and adds a constant to it. This addition is performed over 64-bits.
As an example, the following code snippet increments \texttt{x10} while reading one byte from the memory pointed to by the final value of \texttt{x10}:
\begin{lstlisting}
	ldrb    w18, [x10], #100
	ldrb    w18, [x10], #54
	ldrb    w18, [x10], #-153
\end{lstlisting}

The same limitations apply to \texttt{strb}.

\subsection{Branch operations}
Given the severe restrictions on the minimum offset we can use for branching instructions, only \texttt{tbz} and \texttt{tbnz} instructions are included in $\mathcal A_\text{max}$.

The \texttt{tbz} (test and branch if zero) is given three operands: a bit $b$, a register \texttt{Rt} and a 14 bits immediate value \texttt{imm14}. If the $b^\text{th}$ bit of register \texttt{Rt} is equal to zero, then \texttt{tbz} jumps to the relative offset \texttt{imm14}. 

To keep our shellcode sort, we chose the smallest offset value available, at the expense of restricting our choice for \texttt{Rt} and $b$. We can turn \texttt{tbz} into an unconditional jump by using a register that has been set to zero.
Conditional jumps require the control over a specific register bit, which is trickier. 

The smallest forward alphanumeric jump offset available is 1540 bytes, and the smallest backward jump offset is 4276 bytes. The maximal offset reachable with any of these instructions is less than 1\,MB. \texttt{tbnz} works symmetrically and jumps if the tested bit equals 1.

\section{Fully Alphanumeric AArch64}
\label{sec:A64toAA64}

The building blocks we described so far could be used to assemble complex programs in a bottom-up fashion. However, even though many building blocks could be designed in theory, in practice we get quickly limited by our ability to use branches, system instructions and function calls: Turing-completeness is not enough.

We circumvent this limitation by using a two-stage shellcode, the first being an in-memory alphanumeric decoder (called the \emph{vector}) leveraging the higher-level constructs of the previous section, and the second being our payload $P$ encoded as an alphanumeric string. 

The encoder $\mathcal E$ is written in PHP, while the corresponding decoder $\mathcal D$ is implemented as part of the vector with instructions from $\mathcal A_\text{max}$. Finally, we implemented a linker $L_\mathcal{D}$ that embeds the encoded payload in $\mathcal D$. This operation yields an alphanumeric program $A \gets L_\mathcal{D}(\mathcal E(P))$.

\subsection{The Encoder}
Since we have $62$ alphanumeric characters, it is theoretically possible to encode almost $6$ bits per alphanumeric byte. However, to keep $\mathcal D$ short, we only encode $4$ bits per alphanumeric byte. This spreads each binary byte of the payload $P$ over $2$ alphanumeric consecutive characters. The encoder $\mathcal E$, whose source code can be found in Appendix~\ref{app:program2a}, splits the input byte $P[i]$ and adds \hex{40} to each nibble:
\begin{align*}
a[2i] & \gets (b[i] ~ \& ~ \hex{F}) + \hex{40} \\
a[2i+1] & \gets (b[i] \gg 4) + \hex{40}
\end{align*}
Zero is encoded in a special way: the above encoding would give \hex{40} \emph{i.e.} the character \texttt{`@'}, which does not belong to our alphanumeric character set. We add \hex{10} to the previously computed $a[k]$ to transform it into a \hex{50} which corresponds to \texttt{`P'}.

\subsection{The Decoder}
Some tricks must be used for decoding, as $\mathcal D$ must be an alphanumeric program. Our solution is to use the following snippet:
\begin{lstlisting}
    /* Input: A and B. Z is 0. Output: B */
    eon     wA, wZ, wA, lsl #20
    ands    wB, wB, #0xFFFF000F
    eon     wB, wB, wA, lsr #16
\end{lstlisting}
The first \texttt{eon} shifts \texttt{wA} $20$ bits to the left and negates it, since \texttt{wZ} is zero: 
\begin{align*}
\texttt{wA}_2 \gets \texttt{wZ} \oplus \neg (\texttt{wA}_1 \ll 20) &= \neg (\texttt{wA}_1 \ll 20)
\end{align*}
The \texttt{ands} is used to keep only the $4$ LSBs of \texttt{wB}. The reason why the pattern \hex{FFFF000F} is used (rather than the straightforward \hex{F}) is that the instruction \texttt{ands wB, wB, \hex{FFFF000F}} is alphanumeric, while \texttt{ands wB, wB, \hex{F}} is not. 

The last \texttt{eon} xors $\texttt{wB}$ with the negation of \texttt{wA} shifted $16$ bits to the right, thus recovering the $4$ upper bits.
\begin{align*} 
\texttt{wB} &\gets \texttt{wB} \oplus \neg (\texttt{wA}_2 \gg 16) \\
&= \texttt{wB} \oplus \neg (\neg (\texttt{wA}_1 \ll 20) \gg 16) \\
&= \texttt{wB} \oplus (\texttt{wA}_1 \ll 4)
\end{align*}

It is natural to wish $\mathcal D$ to be as small as possible. However, given that the smallest backward jump requires an offset of 4276 bytes, $\mathcal D$ cannot possibly be smaller than 4276 bytes. 

\subsection{Payload Delivery}

\begin{figure}
    \centering
    \includegraphics[height=.30\textheight]{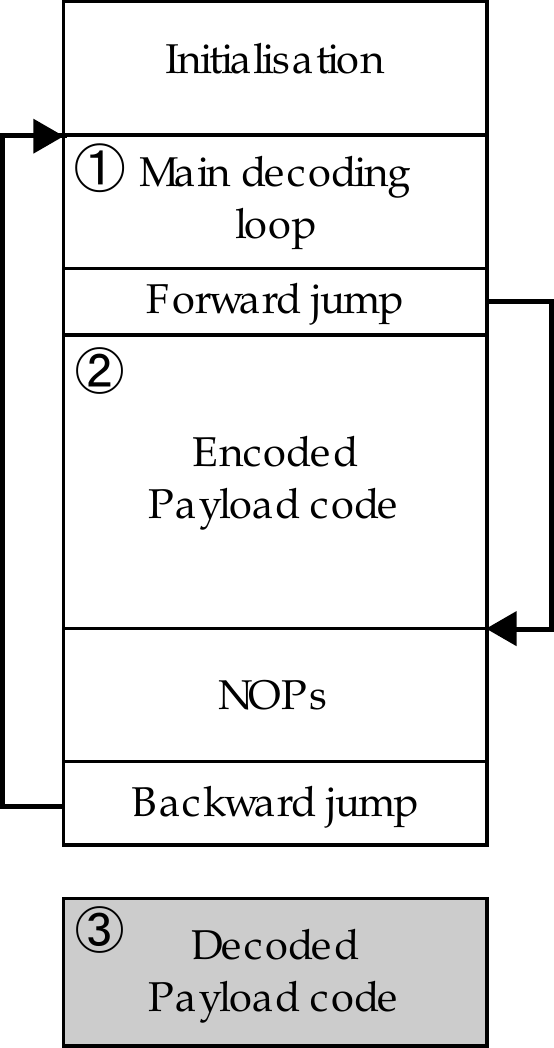}
    \caption{First step: The encoded payload is decoded and placed further down on the stack. Note that (2) is twice the size of (3).}
    \label{fig:stack1}
\end{figure}

\begin{figure}
    \centering
    \includegraphics[height=.30\textheight]{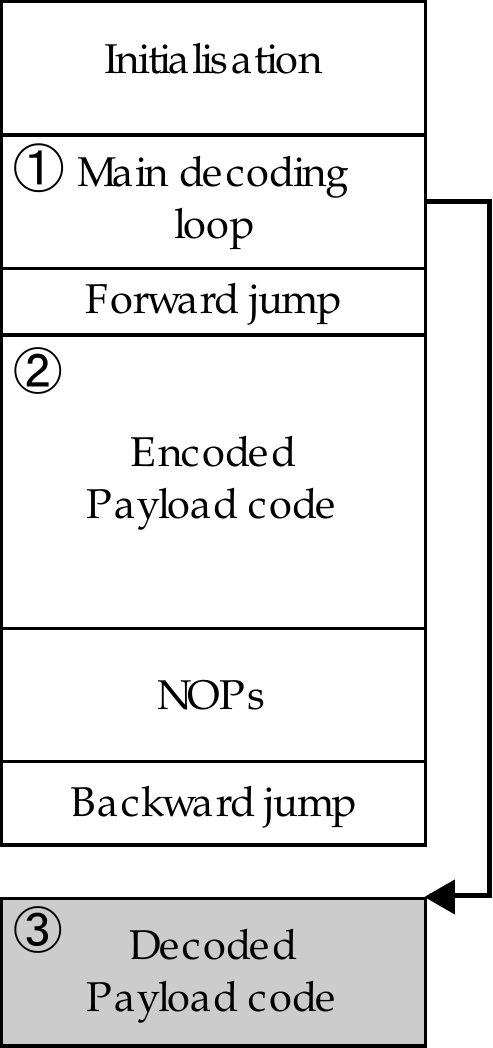}
    \caption{Second step: Once the payload is decoded, the decoder calls it.}
    \label{fig:stack2}
\end{figure}

The encoded payload is embedded directly in $\mathcal D$'s main loop. 
$\mathcal D$ will decode the encoded payload until completion (\emph{cf.} \Cref{fig:stack1}), and will then jumps into the decoded payload (\emph{cf.} \Cref{fig:stack2}).

To implement the main loop we need two jump offsets: one forward offset large enough to jump over the encoded payload, and one even larger backward offset to return to the decoding loop. The smallest available backward offset satisfying these constraints is selected, alongside with the largest forward offset smaller than the chosen backward offset. Extra space is padded with \texttt{nop}-like instructions.

The decoder's source code is provided in Appendix~\ref{app:program2}.

\subsection{Assembly and machine code}
Note that there is no bijection between machine code and assembly.
As an example, \hex{72304F39} (\texttt{9O0r}) is disassembled as 
\begin{lstlisting}
    ands W25, W25, #0xFFFF000F
\end{lstlisting}
but this very instruction, when assembled back, gives \hex{72104F39} (\texttt{9O.r}), which is not alphanumeric. Structurally, \texttt{9O0r} and \texttt{9O.r} are equivalent. However, only the latter is chosen by the assembler.
Thus, to ensure that our generated code is indeed alphanumeric we had to put directly this instruction's hexadecimal representation in the assembly code. 

\subsection {Polymorphic shellcode}
It is possible to add partial polymorphism to both the vector and the payload using our approach. Here our shellcode bypasses basic pattern matching detection methods \cite{Bontchev94} but more specific techniques can be used in order to fool more recent IDS \cite{PhrackPOLY}.

The payload can be mutated using the fact that only the 4 LSBs of each byte contain information about the payload, granting us the possibility to modify the first 4 bits arbitrarily, as long as the instructions still remain alphanumeric. This gives a total polymorphism of the payload as shown by the polymorphic engine provided Appendix~\ref{app:program4}, which mutates each byte into between two and five possibilities. Moreover, the padding following the payload is also mutated with the same code. The NOP sled is also be made totally polymorphic. Indeed, a trivial search reveals more than 80 thousands instructions that could be used as NOP instructions in our shellcode.

The vector is made partially polymorphic by creating different versions of each high level construct. The two easiest ones being the ones defined in sections \ref{sec:zeroingregister} and \ref{sec:incrdecr}, which have both been implemented. Indeed, in order to zero a register, it is possible replace the shift value by anything in the set $\lbrace 16..30 \rbrace \setminus \lbrace 23 \rbrace$. The same idea can be applied to increasing or decreasing a register, in which the immediate value can be replaced by any other constant keeping the instruction alphanumeric (the values are in the range \hex{c0c} - \hex{e5c}, with some gaps in between). We show as an example a polymorphic engine that mutates the zeroing of a register in appendix \ref{app:program4}. Those two techniques are enough to mutate 9 over 25 instructions of the decoder. All in all, we mutated 4256 over 4320 bytes of the shellcode.

\section{Experimental results}

On ARM architectures, when memory is overwritten, the I-cache is not invalidated. This hampers the execution of
self-rewriting code, and has to be circumvented: we need to flush the I-cache for our shellcode to work.
Unfortunately the dedicated instruction to do that is not alphanumeric\footnote{Alternatively, we could assume we were working on a Linux OS and perform the appropriate syscall, but again this instruction is not alphanumeric.}. More precisely, there are only two situations where this is an issue:
\begin{itemize}
	\item Execution of the decoder;
	\item Jump to the decoded payload.
\end{itemize}
Our concern mostly lies with the second point. Fortunately, it is sufficient that the first instructions be not in the cache to invalidate it and flush it with the first instructions. However, cache management is implementation-dependent when it comes to details, making our code less portable. In practice, as the L1 cache is split into a data L1 and a instruction L1, we never ran into a cache coherency issue.

\subsection{QEMU}

As a proof-of-concept, we tested the code with QEMU \cite{qemu}, disregarding the above discussion on cache issues. Moreover, as addresses are below the 4\,GB barrier, we can easily perform pointer arithmetic. We provide in Appendix~\ref{app:program3} the output of our tool, where the input is a simple program printing ``Hello World!''. The result can be easily tested using the parameters given in Appendix~\ref{app:program3}.

\subsection{DragonBoard 410c}

We then moved to real hardware. The DragonBoard 410c \cite{dragonboard} is an AArch64-based board with a Snapdragon 410 SoC. This SoC contains an ARM Cortex A53 64-bit processor. This processor is widely used (in the Raspberry Pi 3 among many others) and is thus representative of the AArch64 world.

We installed Debian 8.0 (Jessie) and were successfully able to run a version of our shellcode.

We had no issue with the I-cache: As we do not execute code on the same page we write, the cache handler does not predict we are going to branch there.

\subsection{Apple iPhone}

Finally we focused on the Apple iPhone\,6 running iOS\,8. Most iOS\,8 applications are developed in the memory-unsafe Objective-C language, and recent research seems to indicate the pervasiveness of vulnerabilities \cite{xing2015unauthorized, PhrackObjC}, all the more since a unicode exploit on CoreText\footnote{Also know as the 'effective power' SMS exploit} working on early iOS\,8 has been released, which consists in a corruption of a pointer being then dereferenced.

We build an iPhone application to test our approach. For the sake of credibility, we shaped our scenario on existing applications that are currently available on the Apple Store.
Thus, although we made the application vulnerable on purpose, we stress that such a vulnerability
could realistically be found in the wild.

Namely, the scenario is as follows:
\begin{itemize}
\item The application loads some \emph{statically} compiled scripts, which are based on players parameters
\item It also \emph{interprets} the downloaded scripts (they cannot be compiled per Apple guidelines)
\item Downloaded scripts (for example scripts made by users) are sanity-checked (must be printable characters: blanks + \hex{20}-\hex{7E} range)
\item Thus, there is an array of tuples $\{t, p\}$ in which $t$ indicates interpreted script or JIT compiled executable code, and $p$ is the pointer to the aforementioned script or code.
\item A subtle bug enables an attacker to assign the \emph{wrong} type of script in certain cases
\item Thus we can force our ill-intentioned user-script to be considered as executable code instead of interpretable script.
\item Therefore our shellcode gets called as a function directly.
\end{itemize}

From then on, the decoder retrieves the payload and uses a gadget to change the page permissions from ``write'' to ``read|exec''\footnote{Apple iOS enforces write xor exec.}, and executes it.

In this proof-of-concept, our shellcode only changes the return value of a function, displaying an incorrect string on the screen.

\section{Conclusion}

We described a methodology as well as a generic framework to turn arbitrary code into an (equivalent) executable alphanumeric program for ARMv8 platforms. To the best of our knowledge, no such tools are available for this platform, and up to this point most constructions were only theoretical.

Our final construction relies on a fine-grained understanding of ARMv8 specifics, yet the overall strategy is not restricted to that
processor, and may certainly be transposed to address other architectures and constraints. 

\bibliographystyle{abbrv}
\bibliography{ref}

\begin{thebibliography}{10}

\bibitem{PhrackBOF}
Aleph-One.
\newblock Smashing the stack for fun and profit.
\newblock {\em Phrack}, 49, 1996.
\newblock {\small\url{http://phrack.org/issues/49/14.html}}.

\bibitem{ARMv8Ap536}
ARM Limited, 110 Fulbourn Road, Cambridge.
\newblock {\em ARM Architecture Reference Manual. ARMv8, for ARMv8-A
  architecture profile}, 2013.
\newblock
  {\small\url{https://static.docs.arm.com/ddi0487/db/DDI0487D_b_armv8_arm.pdf}}.

\bibitem{Basu}
A.~Basu, A.~Mathuria, and N.~Chowdary.
\newblock {Automatic Generation of Compact Alphanumeric Shellcodes for x86}.
\newblock In {\em Proceedings of the 10th International Conference on
  Information Systems Security}, pages 399--410, Berlin, Heidelberg, 2014.
  Springer-Verlag.
\newblock {\small\url{https://doi.org/10.1007/978-3-319-13841-1\_22}}.

\bibitem{qemu}
F.~Bellard.
\newblock {QEMU, a Fast and Portable Dynamic Translator}.
\newblock In {\em Proceedings of the 2005 USENIX Annual Technical Conference},
  pages 41--46, Berkeley, CA, USA, 2005. USENIX Association.
\newblock
  {\small\url{https://www.cse.iitd.ernet.in/~sbansal/csl862-virt/2010/readings/bellard.pdf}}.

\bibitem{Bontchev94}
V.~Bontchev.
\newblock Future trends in virus writing.
\newblock {\em International Review of Law, Computers \& Technology},
  11(1):129--146, 1997.
\newblock {\small\url{https://bontchev.nlcv.bas.bg/papers/docs/trends.txt}}.

\bibitem{cristofani07}
D.~B. Cristofani.
\newblock A universal {T}uring machine, 1996.
\newblock {\small\url{http://www.hevanet.com/cristofd/brainfuck/utm.b}}.

\bibitem{davi2011privilege}
L.~Davi, A.~Dmitrienko, A.-R. Sadeghi, and M.~Winandy.
\newblock Privilege escalation attacks on {A}ndroid.
\newblock In {\em Proceedings of the 13th International Conference on
  Information Security}, pages 346--360, Berlin, Heidelberg, 2011.
  Springer-Verlag.
\newblock
  {\small\url{https://www.researchgate.net/publication/220905164_Privilege_Escalation_Attacks_on_Android}}.

\bibitem{PhrackPOLY}
T.~Detristan, T.~Ulenspiegel, Y.~Malcom, and M.~S. vonUnderduk.
\newblock Polymorphic shellcode engine using spectrum analysis.
\newblock {\em Phrack}, 61, 2003.
\newblock {\small\url{http://phrack.org/issues/61/9.html}}.

\bibitem{Eller}
R.~Eller.
\newblock Bypassing {MSB} data filters for buffer overflow exploits on {I}ntel
  platforms, 2000.
\newblock {\small\url{https://securiteam.com/securityreviews/5dp140afpq/}}.

\bibitem{faase07}
Faase.
\newblock Bf is {T}uring-complete.
\newblock {\small\url{http://www.iwriteiam.nl/Ha_bf_Turing.html}}.

\bibitem{kernighan1977m4}
B.~W. Kernighan and D.~M. Ritchie.
\newblock {\em The {M4} macro processor}.
\newblock Bell Laboratories, 1977.
\newblock
  {\small\url{https://wolfram.schneider.org/bsd/7thEdManVol2/m4/m4.pdf}}.

\bibitem{DBLP:conf/ccs/MasonSMM09}
J.~Mason, S.~Small, F.~Monrose, and G.~MacManus.
\newblock {English Shellcode}.
\newblock In {\em Proceedings of the 16th {ACM} Conference on Computer and
  Communications Security}, pages 524--533, New York, NY, 2009. ACM.
\newblock {\small\url{https://web.cs.jhu.edu/~sam/ccs243-mason.pdf}}.

\bibitem{Metasploit}
{Metasploit Project}.
\newblock The {M}etasploit {F}ramework.
\newblock {\small\url{http://www.metasploit.com/}}.

\bibitem{PhrackObjC}
Nemo.
\newblock Modern {Objective-C Exploitation Techniques}.
\newblock {\em Phrack}, 69, 2016.
\newblock {\small\url{http://phrack.org/issues/69/9.html}}.

\bibitem{PhrackUnicode}
Obscou.
\newblock Building {IA32} {U}nicode-proof shellcodes.
\newblock {\em Phrack}, 61, 2003.
\newblock {\small\url{http://phrack.org/issues/61/11.html}}.

\bibitem{dragonboard}
Qualcomm.
\newblock Dragonboard 410c.
\newblock
  {\small\url{https://developer.qualcomm.com/hardware/dragonboard-410c}}.

\bibitem{raiter07}
B.~Raiter.
\newblock Brainfuck.
\newblock {\small\url{http://www.muppetlabs.com/~breadbox/bf/}}.

\bibitem{PhrackIA32}
RIX.
\newblock Writing {IA32} alphanumeric shellcodes.
\newblock {\em Phrack}, 57, 2001.
\newblock {\small\url{http://phrack.org/issues/57/15.html}}.

\bibitem{tan2008empirical}
G.~Tan and J.~Croft.
\newblock An empirical security study of the native code in the jdk.
\newblock In {\em Proceedings of the 17th Conference on Security Symposium},
  pages 365--378, Berkeley, CA, USA, 2008. USENIX Association.
\newblock
  {\small\url{https://www.usenix.org/legacy/event/sec08/tech/full_papers/tan_g/tan_g.pdf}}.

\bibitem{xing2015unauthorized}
L.~Xing, X.~Bai, T.~Li, X.~Wang, K.~Chen, X.~Liao, S.-M. Hu, and X.~Han.
\newblock Cracking app isolation on apple: Unauthorized cross-app resource
  access on mac os~x and ios.
\newblock In {\em Proceedings of the 22nd ACM SIGSAC Conference on Computer and
  Communications Security}, pages 31--43, New York, NY, USA, 2015. ACM.
\newblock
  {\small\url{https://www.informatics.indiana.edu/xw7/papers/xing2015cracking.pdf}}.

\bibitem{Younan}
Y.~Younan, P.~Philippaerts, F.~Piessens, W.~Joosen, S.~Lachmund, and T.~Walter.
\newblock Filter-resistant code injection on {ARM}.
\newblock {\em Journal in Computer Virology}, 7(3):173--188, 2011.
\newblock
  {\small\url{http://amnesia.gtisc.gatech.edu/~moyix/CCS_09/docs/p11.pdf}}.

\end{thebibliography}

\appendix

\section{Source code of Program 1}
\label{app:program1}

The following Haskell program generates all possible combinations of 4 alphanumeric characters,
and writes them on the disk.

\begin{lstlisting}[language=Haskell]
    n = [[a,b,c,d]|a<-i,b<-i,c<-i,d<-i] 
        where i = ['0'..'9']++
                  ['a'..'z']++
                  ['A'..'Z']
    
    m = concat n
    main = writeFile "allalphanum" m
\end{lstlisting}
    
\section{Alphanumeric Instructions}
\label{app:list1}
This appendix describes $\mathcal A_1$, the set of all AArch64 opcodes that can give alphanumeric instructions 
for some operands.
\begin{itemize}
\item Data processing instructions:
\begin{lstlisting}[xleftmargin=0cm]
adds, sub, subs, adr, bics, ands, orr, eor, eon, ccmp
\end{lstlisting}
\item Load and store instructions
\begin{lstlisting}[xleftmargin=0cm]
ldr, ldrb, ldpsw, ldnp, ldp, ldrh, ldurb, ldxrh, ldtrb, ldtrh, ldurh, strb, stnp, stp, strh
\end{lstlisting}
\item Branch instructions
\begin{lstlisting}[xleftmargin=0cm]
cbz, cbnz, tbz, tbnz, b.cond
\end{lstlisting}
\item Other (SIMD, floating point, crypto...)
\begin{lstlisting}[xleftmargin=0cm]
cmhi, shl, cmgt, umin, smin, smax, umax, usubw2, ushl, srshl, sqshl, urshl, uqshl, sshl, ssubw2, rsubhn2, sqdmlal2, subhn2, umlsl2, smlsl2, uabdl2, sabdl2, sqdmlsl2, fcvtxn2, fcvtn2, raddhn2, addhn2, fcvtl2, uqxtn2, sqxtn2, uabal2, sabal2, sri, sli, uabd, sabd, ursra, srsra, uaddlv, saddlv, sqshlu, shll2, zip2, zip1, uzp2, mls, trn2
\end{lstlisting}
\end{itemize}

\section{Alphanumeric AND}
\label{app:codeAND}

The \texttt{and} operation described in \Cref{sec:AND} can be automatically generated using the
following code.
To abstract register numbers and generate repetitive lines, the source code provided is
pre-processed by \texttt{m4} \cite{kernighan1977m4}. This allowed us to easily change a register number without changing 
every occurrence if we found that a specific register could not be used.

\begin{lstlisting}
divert(-1)
changequote({,})
define({LQ},{changequote(`,'){dnl}
changequote({,})})
define({RQ},{changequote(`,')dnl{
}changequote({,})})
changecom({;})

define({concat},{$1$2})dnl
define({A}, 18)
define({B}, 25)
define({C}, 17)
define({D}, 11)
define({E}, 19)
define({F}, 26)
define({WA}, concat(W,A))
define({WB}, concat(W,B))
define({WC}, concat(W,C))
define({WD}, concat(W,D))
define({WE}, concat(W,E))
define({WF}, concat(W,F))

divert(0)dnl

ands WD, WD, WD, lsr #16
ands WD, WD, WD, lsr #16
ands WC, WC, WC, lsr #16
ands WC, WC, WC, lsr #16
ands WE, WE, WE, lsr #16
ands WE, WE, WE, lsr #16
ands WF, WF, WF, lsr #16
ands WF, WF, WF, lsr #16
eon  WC, WC, WB, lsl #16
eon  WE, WE, WA, lsl #16
eon  WF, WF, WE, lsr #16
bics WD, WF, WC, lsr #16
ands WC, WC, WC, lsr #16
ands WC, WC, WC, lsr #16
ands WE, WE, WE, lsr #16
ands WE, WE, WE, lsr #16
ands WF, WF, WF, lsr #16
ands WF, WF, WF, lsr #16
eon  WC, WC, WB, lsr #16
eon  WE, WE, WA, lsr #16
eon  WF, WF, WE, lsl #16
bics WD, WF, WC, lsl #16
\end{lstlisting}

\section{Encoder's Source Code}
\label{app:program2a}
We give here the encoder's full source code. This program is written in PHP.

\begin{lstlisting}[language=php]
function mkchr($c) {
	return(chr(0x40 + $c));
}

$s = file_get_contents('shellcode.bin.tmp');
$p = file_get_contents('payload.bin');
$b = 0x60;  /* Synchronize with pool */
for($i=0; $i<strlen($p); $i++)
{
	$q = ord($p[$i]);
	$s[$b+2*$i  ] = mkchr(($q >> 4) & 0xF);
	$s[$b+2*$i+1] = mkchr( $q       & 0xF);
}
$s = str_replace('@', 'P', $s);
file_put_contents('shellcode.bin', $s);
\end{lstlisting}

\section{Decoder's Source Code}
\label{app:program2}
We give here the decoder's full source code. This code is pre-processed by \texttt{m4} \cite{kernighan1977m4} which performs macro expansion. The payload has to be be placed at the offset designated by the label \texttt{pool}.

\begin{lstlisting}
divert(-1)
changequote({,})
define({LQ},{changequote(`,'){dnl}
changequote({,})})
define({RQ},{changequote(`,')dnl{
}changequote({,})})
changecom({;})

define({concat},{$1$2})dnl
define({repeat}, {ifelse($1, 0, {}, $1, 1, {$2}, {$2
        repeat(eval($1-1), {$2})})})

define({P}, 10)
define({Q}, 11)
define({S},  2)
define({A}, 18)
define({B}, 25)
define({U}, 26)
define({Z}, 19)

define({WA}, concat(W,A))
define({WB}, concat(W,B))
define({WP}, concat(W,P))
define({XP}, concat(X,P))
define({WQ}, concat(W,Q))
define({XQ}, concat(X,Q))
define({WS}, concat(W,S))
define({WU}, concat(W,U))
define({WZ}, concat(W,Z))
divert(0)dnl

/* Set P */
l1:     ADR     XP, l1+0b010011000110100101101
        /* Sync with pool */
        SUBS    WP, WP, #0x98, lsl #12
        SUBS    WP, WP, #0xD19

/* Set Q */
l2:     ADR     XQ, l2+0b010011000110001001001
        /* Sync with TBNZ */
        SUBS    WQ, WQ, #0x98, lsl #12
        ADDS    WQ, WQ, #0xE53
        ADDS    WQ, WQ, #0xC8C

/* Z:=0 */
        ANDS    WZ, WZ, WZ, lsr #16
        ANDS    WZ, WZ, WZ, lsr #16

/* S:=0 */
        ANDS    WS, WZ, WZ, lsr #12

/* Branch to code */
loop:   TBNZ    WS, #0b01011, 0b0010011100001100

/* Load first byte in A */
        LDRB    WA, [XP, #76]
/* Load second byte in B */
        LDRB    WB, [XP, #77]
/* P+=2 */
        ADDS    WP, WP, #0xC1B
        SUBS    WP, WP, #0xC19

/* Mix A and B */
        EON     WA, WZ, WA, lsl #20
        /* ANDS WB, WB, #0xFFFF000F */
        .word   0x72304C00+33*B
        EON     WB, WB, WA, lsr #16

/* STRB B, [Q] */
        STRB    WB, [XQ, WZ, uxtw]

/* Q++ */
        ADDS    WQ, WQ, #0xC1A
        SUBS    WQ, WQ, #0xC19

/* S++ */
        ADDS    WS, WS, #0xC1A
        SUBS    WS, WS, #0xC19

        TBZ     WZ, #0b01001, next

pool:   repeat(978, {.word 0x42424242})

/* NOPs */
next:   repeat( 77, {ANDS WU, WU, WU, lsr #12})

        TBZ     WZ, #0b01001, loop
\end{lstlisting}
    
    \pagebreak
\section{Hello World Shellcode}
\label{app:program3}    

The following program prints ``Hello world'' when executed in QEMU (tested with \texttt{qemu-system-aarch64 -machine virt -cpu cortex-a57 -machine type=virt -nographic -smp 1 -m 2048 -kernel shellcode.bin  --append "console=ttyAMA0"}).
It was generated by the program described in \Cref{sec:A64toAA64}.

The notation \texttt{(X)\^{}\{Y\}} means that \texttt{X} is
repeated \texttt{Y} times.    
\begin{lstlisting}
jiL0JaBqJe4qKbL0kaBqkM91k121sBSjsBSjb2Sj
b8Y7R1A9Y5A9Jm01Je0qrR2J9O0r9CrJyI38ki01
ke0qBh01Bd0qszH6PPBPJHMBAOPPPPIAAKPPPPID
PPPPPPADPPALPPECPBBPJAMBPAPCHPMBPABPJAOB
BAPPDPOIJAOOBOCGPAALPPECAOBHPPGADAPPPPOI
FAPPPPEDJPPAHPEBOGOOOOAGLPPCEOMFOMGKKNJI
OMPCPPIAOCPKPPOIOCPCPPJJFPPBDPCIHPPPPPCD
GCPFPPIANLOOOOIGOLOOOOAGOCPKDPOIOMGKLBJH
LPPCEOMFOMGKKOJIPPPMHPEBOMPCPPIANDOOOOIG
JPPLHPEBNBOOOOIGHPPMHPEBNPOOOOIGHPPMHPEB
MNOOOOIGNPPMHPEBMLOOOOIGHPPEHPEBMJOOOOIG
PPPDHPEBMHOOOOIGNPPNHPEBMFOOOOIGNPPMHPEB
MDOOOOIGDPPNHPEBMBOOOOIGHPPMHPEBMPOOOOIG
HPPLHPEBLNOOOOIGBPPDHPEBLLOOOOIGDPPAHPEB
LJOOOOIGPPPPHPEBOMGKLAJHLPPCEOMF
(BBBB)^{854}
(Z3Zj)^{77}
szO6
\end{lstlisting}
\section{Polymorphic engine}
\label{app:program4}
The following shows two modifications that make the code partly polymorphic. The first one is a modification of the encoder, that will randomize both the payload and the remaining blank space.

\begin{lstlisting}[language=php]
function mkchr($c) {
	$a = [];
	if($c>0x0){ $a[] = 0x40; $a[] = 0x60;}
	if($c<0xA){ $a[] = 0x30;}
	if($c<0xB){ $a[] = 0x50; $a[] = 0x70;}
	return(chr($a[array_rand($a)]+$c));
}
	
function randalnum() {
	$n = rand(0, 26+26+10-1);
	if($n<26) {	return chr(0x41 + $n); }
	$n -= 26;
	if($n<26) { return chr(0x61 + $n); }
	return chr(0x30 + $n - 26);
}

/* Replace '$s = str_replace('@', 'P', $s);' with: */
$j = $b + 2*$i;
while($s[$j] === 'B') {
	$s[$j++] = randalnum();
}
\end{lstlisting}

The second one is an example of adding polymorphism for zeroing a register using a Haskell engine. 

\begin{lstlisting}[language=Haskell]
import Data.String.Utils
import Data.List
import Data.Random
                
shift = "SHIFT"
shiftRange = [16..22]++[24..30]

replacePoly :: String -> String -> RVar String
replacePoly acc [] = return $ reverse acc
replacePoly acc s = do
  if (startswith shift s)
  then do
    randomSh <- randomElement shiftRange
    replacePoly ((reverse $ "#" ++ (show randomSh))++acc)
      $ drop (length shift) s
  else do
    replacePoly ((head s):acc) $ tail s
  
main = do
  s <- readFile "vector.a64"
  sr <- runRVar (replacePoly [] s) StdRandom
  writeFile "vector.a64.poly" sr

\end{lstlisting}

\end{document}